\documentclass[english]{article}
\usepackage[T1]{fontenc}
\usepackage[latin9]{inputenc}
\usepackage{color}
\usepackage{amssymb}

\makeatletter
\newcommand{\lyxaddress}[1]{
\par {\raggedright #1
\vspace{1.4em}
\noindent\par}
}

\makeatother

\usepackage{babel}
\begin{document}

\title{\textbf{Space-time can be neither discrete nor continuous}}

\author{\textbf{Christian Corda}}
\maketitle

\lyxaddress{\begin{center}
Research Institute for Astronomy and Astrophysics of Maragha (RIAAM),
P.O. Box 55134-441, Maragha, Iran and International Institute for
Applicable Mathematics and Information Sciences\textit{, }Hyderabad\textit{,
}\textit{\emph{INDIA}}
\par\end{center}}

\begin{center}
\textit{E-mail address:} \textcolor{blue}{cordac.galilei@gmail.com} 
\par\end{center}
\begin{abstract}
We show that our recent Bohr-like approach to black hole (BH) quantum
physics implies that space-time quantization could be energy-dependent.
Thus, in a certain sense, space-time can be neither discrete nor continuous.
Our approach permits also to show that the ``volume quantum'' of
the Schwarzschild space-time increases with increasing energy during
BH evaporation and arrives to a maximum value when the Planck scale
is reached and the generalized uncertainty principle (GUP) prevents
the total BH evaporation. Remarkably, this result does not depend
on the BH original mass. The interesting consequence is that the behavior
of BH evaporation should be the same for all Schwarzschild BHs when
the Planck scale is approached.
\end{abstract}
\begin{quotation}
\emph{To the memory of Stephen W. Hawking.}
\end{quotation}
The search for a theory of quantum gravity (TQG) through BH physics
started in the '70s of last century with the famous papers of Bekenstein
\cite{key-1} and Hawking \cite{key-2}. The famous formula of the
Bekenstein-Hawking entropy \cite{key-1,key-2} 
\begin{equation}
S_{BH}=\frac{c^{3}k_{B}A}{4G\hbar},\label{eq: Bekenstein-Hawking}
\end{equation}
where $S_{BH}$ stands for the BH entropy, $A$ for the BH surface
area (the event horizon), $\hbar$ is the reduced Planck constant,
$c$ is the speed of light, $k_{B}$ is the Boltzmann constant and
$G$ is the gravitational constant, is indeed considered very fundamental.
In fact, on one hand it counts the BH effective degrees of freedom.
On the other hand, it ties together notions from gravitation, thermodynamics
and quantum theory. Hence, it includes all the 3 fundamental constants
in, and it is, in turn, considered an important window into the yet
unknown TQG. In addition, the discovery by Hawking of BH radiation
\cite{key-2} represents the first, non-banal result of combining
Einstein's general relativity with Heisenberg's uncertainty principle.
After those pioneering works, an enormous amount of papers have been
written and currently continue to be written on BH quantum physics.
Today, there is indeed a large agreement on the idea that BHs should
be highly excited states representing the fundamental bricks of the
yet unknown TQG \cite{key-3}. This idea represents a parallelism
with QM. In fact, in the '20s of last century, atoms were considered
the fundamental bricks of QM. This analogy enables one to argue that
the BH mass could have a discrete spectrum \cite{key-3}. On the other
hand, an immediate and natural question surfaces from such a parallelism.
If one assumes that the BH should be the ``nucleus'' of the ``gravitational
atom'', then it is quite natural asking: What are the ``electrons''?
In a series of recent papers (see {[}4 - 6{]} and references within),
an intriguing answer addressed such a question. The BH quasi-normal
modes (QNMs) (which are the horizon's oscillations in a semi-classical
approach \cite{key-4}), which are ``triggered'' by absorptions
of external particles and by the emission of Hawking radiation, represent
the ``electrons'' of that ``gravitational atom''. In fact, in
{[}4 - 6{]} it has been shown that the semi-classical evaporating
Schwarzschild BH is the gravitational analogous of the historical,
semi-classical hydrogen atom, introduced by Niels Bohr in 1913 \cite{key-7,key-8}.
The idea underlying the results in {[}4 - 6{]} is founded on the non-thermal
spectrum of Hawking radiation in \cite{key-9}. This indeed implies
the countable character of subsequent emissions of Hawking quanta,
which, in turn, generates an obvious and important correspondence
between Hawking radiation \cite{key-2} and the BH QNMs {[}4 - 6{]}.
In the framework in {[}4 - 6{]}, QNMs are seen as being the \textquotedbl{}electron
states\textquotedbl{}, jumping from a quantum level to another one.
The analogy is completed when one identifies the energy \textquotedbl{}shells\textquotedbl{}
of the ``gravitational hydrogen atom'' in terms of the absolute
values of the quasi-normal frequencies {[}4 - 6{]}. Another remarkable
result is that the BH information puzzle can be solved by considering
the time evolution of the Bohr-like BH \cite{key-5,key-6}. In fact,
BH evaporation results governed by a time-dependent Schrodinger equation,
while subsequent emissions of Hawking quanta are entangled with the
BH ``electron states'', i.e. with the QNMs \cite{key-5,key-6}.
The results in {[}4 - 6{]} are completely consistent with previous
literature. In particular, the result of Bekenstein on the area quantization
\cite{key-10} is in complete agreement with the works {[}4 - 6{]}.
For the sake of completeness, we stress that our Bohr-like approach
to BH quantum physics has been recently generalized to the Large AdS
BHs in \cite{key-11}. Hereafter, we will use Planck units ($G=c=k_{B}=\hbar=\frac{1}{4\pi\epsilon_{0}}=1$)
for the sake of simplicity. Then, for large values of the principal
quantum number $n$ (i.e. for excited BHs), the energy levels of the
Schwarzschild BH, which is interpreted as the ``gravitational hydrogen
atom'', are given by {[}4 - 6{]} 
\begin{equation}
E_{n}\equiv|\omega_{n}|=M-\sqrt{M^{2}-\frac{n}{2}},\label{eq: radice fisica}
\end{equation}
where $M$ is the initial BH mass and $E_{n}$ represents the total
energy emitted when the BH is excited at the level $n$ {[}4 - 6{]}.
The BH radiates a discrete amount of energy in a quantum jump, and,
for large values of $n,$ the process results independent of the other
quantum numbers. This is in perfect agreement with \emph{the Correspondence
Principle }stated by Bohr in 1920\emph{ }\cite{key-12}.\emph{ }Bohr's
Correspondence Principle argues indeed that \textquotedblleft \emph{transition
frequencies at large quantum numbers should equal classical oscillation
frequencies}\textquotedblright{} \cite{key-12}. In Bohr's 1913 approach
{[}7, 8{]}, electrons only gain and lose energy through quantum jumps
between different allowed energy shells. In each jump, the atom absorbs
or emits radiation with an energy difference between the two involved
levels which is given by the Planck relation (in standard units) $E=hf$,
($\:h\:$ is the Planck constant and $f\:$ the frequency of the involved
transition). In the current approach, the BH QNMs gain and lose energy
through quantum jumps from one allowed energy shell to another with
absorbed or emitted (Hawking) radiation. The energy difference between
the two levels is given by {[}4 - 6{]}
\begin{equation}
\begin{array}{c}
\Delta E_{n_{1}\rightarrow n_{2}}\equiv E_{n_{2}}-E_{n_{1}}=M_{n_{1}}-M_{n_{2}}=\\
\\
=\sqrt{M^{2}-\frac{n_{1}}{2}}-\sqrt{M^{2}-\frac{n_{2}}{2}},
\end{array}\label{eq: jump}
\end{equation}
This equation governs the energy jump between two generic, allowed
levels $n_{1}$ and $n_{2}>n_{1}$. Such a jump is due to the emission
of a particle having frequency $\Delta E_{n_{1}\rightarrow n_{2}}$.
In Eq. (\ref{eq: jump}), $M_{n}$ is the residual mass of the BH
excited at the level $n.$ It is given by the original BH mass minus
the total energy emitted when the BH is excited at that level {[}4
- 6{]}. Hence, $M_{n}=M-E_{n},$ and the jump between the two generic
allowed levels depends only on the initial BH mass and on the two
different values of the BH principal quantum number {[}4 - 6{]}. Instead,
the case of an absorptions is governed by the equation {[}4 - 6{]}
\begin{equation}
\begin{array}{c}
\Delta E_{n_{2}\rightarrow n_{1}}\equiv E_{n_{1}}-E_{n_{2}}=M_{n_{2}}-M_{n_{1}}=\\
\\
=\sqrt{M^{2}-\frac{n_{2}}{2}}-\sqrt{M^{2}-\frac{n_{1}}{2}}=-\Delta E_{n_{1}\rightarrow n_{2}.}
\end{array}\label{eq: jump absorption}
\end{equation}
The analogy with Bohr's hydrogen atom is finalized by the following
intriguing remark. The interpretation of Eq. (\ref{eq: radice fisica})
is of a particle, the ``electron'' of the ``gravitational atom'',
which is quantized on a circle of length {[}4 - 6{]} 
\begin{equation}
L=4\pi\left(M+\sqrt{M^{2}-\frac{n}{2}}\right).\label{eq: lunghezza cerchio}
\end{equation}
This is exactly the analogous of the electron which travels in circular
orbits around the nucleus in Bohr's hydrogen atom {[}7, 8{]}, and
is also similar in structure to the solar system.

For the goals of this paper, the key point is the following. As we
stressed above, in {[}4 - 6{]} we have shown that our results are
in full agreement with the result of Bekenstein on the area quantization
\cite{key-10}. The area of the BH horizon is indeed quantized in
units of the Planck length ($l_{p}=1.616\times10^{-33}\mbox{ }cm$
is equal to one in Planck units) and Bekenstein has shown that the
Schwarzschild BH area quantum is $\triangle A=8\pi$ \cite{key-10}.
The analysis in {[}4 - 6{]} found the same result of Bekenstein for
an energy jump among two allowed neighboring levels $n$ and $n-1$
as 
\begin{equation}
|\triangle A_{n}|=|\triangle A_{n-1}|=8\pi.\label{eq: 8 pi planck}
\end{equation}
Thus, recalling that, in Schwarzschild BHs, the gravitational radius
$r_{g}$ is connected with the BH mass through the relation $r_{g}\equiv2M$
\cite{key-13,key-14}, one defines the gravitational radius associated
to the BH quantum level \emph{$n$} as 
\begin{equation}
r_{g(n)}\equiv2M_{n}.\label{eq: raggio ennesimo}
\end{equation}
From Eq. (\ref{eq: jump}), one sees immediately that the variation
of the gravitational radius due to an emission from the two levels
$n_{1}$ and $n_{2}>n_{1}$ is 
\begin{equation}
\begin{array}{c}
\Delta r_{g\left(n_{1}\rightarrow n_{2}\right)}\equiv r_{g(n_{1})}-r_{g(n_{2})}=2\left(E_{n_{2}}-E_{n_{1}}\right)=\\
\\
=2\left(M_{n_{1}}-M_{n_{2}}\right)=2\left(\sqrt{M^{2}-\frac{n_{1}}{2}}-\sqrt{M^{2}-\frac{n_{2}}{2}}\right),
\end{array}\label{eq: jump 2}
\end{equation}
while the variation of the gravitational radius due to an absorption
from the two levels $n_{2}$ and $n_{1}$ is 
\begin{equation}
\begin{array}{c}
\Delta r_{g\left(n_{2}\rightarrow n_{1}\right)}\equiv r_{g(n_{2})}-r_{g(n_{1})}=2\left(E_{n_{1}}-E_{n_{2}}\right)=\\
\\
=2\left(M_{n_{2}}-M_{n_{1}}\right)=2\left(\sqrt{M^{2}-\frac{n_{2}}{2}}-\sqrt{M^{2}-\frac{n_{1}}{2}}\right)=-\Delta r_{g\left(n_{1}\rightarrow n_{2}\right)}.
\end{array}\label{eq: jump 2 absorption}
\end{equation}
Then, using Eqs. (\ref{eq: 8 pi planck}) and (\ref{eq: jump 2}),
or Eqs. (\ref{eq: 8 pi planck}) and (\ref{eq: jump 2 absorption}),
one finds immediately that the variation of the Schwarzschild ``volume
quantum'', corresponding to a transition between two neighboring
levels $n$ and $n-1$, is 
\begin{equation}
\begin{array}{c}
\Delta V_{\left(n-1\rightarrow n\right)}=\Delta V_{\left(n\rightarrow n-1\right)}\equiv|\Delta r_{g\left(n-1\rightarrow n\right)}||\triangle A_{n}|=\\
\\
=|\Delta r_{g\left(n\rightarrow n-1\right)}||\triangle A_{n-1}|=16\pi|\left(\sqrt{M^{2}-\frac{n}{2}}-\sqrt{M^{2}-\frac{n-1}{2}}\right)|.
\end{array}\label{eq: volume quantum}
\end{equation}
We wrote ``volume quantum'' within inverted commas because the difference
between two Schwarzschild radial coordinates is less than the correspondent
physical proper distance \cite{key-13,key-14}. We also recall that
the Schwarzschild radial coordinate is space-like and time-like outside
and inside the BH event horizon, respectively, and that the horizon
area is a proper area \cite{key-14}. 

Eq. (\ref{eq: volume quantum}) is intriguing because it shows that
the variation of the Schwarzschild ``spatial volume'' of the external
BH space-time, due to an emission/absorption of a particle, is not
constant, but depends on the BH principal quantum number $n$ (that
means on the BH energy level), and on the BH initial mass. On the
other hand, we recall that the singularity of the Schwarzschild radius
is not a real physical singularity \cite{key-14}. It is a coordinate
singularity instead \cite{key-14}. In other words, the space-time
geometry is well behaved at the Schwarzschild radius \cite{key-14}.
This means that, if the variation of the Schwarzschild ``spatial
volume'' is energy dependent, we can reasonably argue that also the
variation of the proper spatial volume is energy dependent. Thus,
space-time quantization seems to be energy dependent. In other words,
space-time can be considered as being neither discrete nor continuous. 

Now, one recalls that BHs cannot emit more energy than their total
mass. In addition, the total energy emitted by the BH cannot be imaginary
{[}4 - 6{]}. Consequently, it must exist a maximum value of the principal
quantum number $n$ {[}4 -6{]}. On the other hand, in \cite{key-15}
it has been shown that the GUP prevents the total BH evaporation in
analogous way that the uncertainty principle prevents the hydrogen
atom from the total collapse. In fact, the collapse is prevented by
dynamics, rather than by symmetry, when the \emph{Planck scale} is
approached \cite{key-15}. This fixes the maximum value of the principal
quantum number $n$ as {[}4 - 6{]}

\begin{equation}
n_{max}=2(M^{2}-1).\label{eq: n max 1}
\end{equation}
Let us compute the prime derivative of $\Delta r_{g\left(n-1\rightarrow n\right)}$
(that is the variation of the Schwarzschild radius due to an emission
between two neighboring levels $n$ and $n-1$ ) with respect to $n$.
One obtains
\begin{equation}
\frac{d}{dn}\left[\Delta r_{g\left(n-1\rightarrow n\right)}\right]=4\pi\left(\frac{\sqrt{M^{2}-\frac{n-1}{2}}-\sqrt{M^{2}-\frac{n}{2}}}{\sqrt{\left(M^{2}-\frac{n}{2}\right)\left(M^{2}-\frac{n-1}{2}\right)}}\right).\label{eq: prime derivative}
\end{equation}
$\frac{d}{dn}\left[\Delta r_{g\left(n-1\rightarrow n\right)}\right]$
is not defined when $n=2M^{2}$ and $n=2M^{2}+1$ which are values
of $n$ forbidden by the constraint (\ref{eq: n max 1}). Analysing
the sign of $\frac{d}{dn}\left[\Delta r_{g\left(n-1\rightarrow n\right)}\right]$
one sees that it always positive for the permitted values of $n,$
which are $1\ll n\leq n_{max}.$ Thus, as $\Delta r_{g\left(n-1\rightarrow n\right)}$
is always positive (considering emissions, the Schwarzschild radius
decreases, but we defined $\Delta r_{g\left(n_{1}\rightarrow n_{2}\right)}$
in Eq. (\ref{eq: jump 2}) as the difference between a longer and
a smaller Schwarzschild radius in case of emissions), its value increases
with increasing $n$, and $n_{max}$ is a maximum for $\Delta r_{g\left(n-1\rightarrow n\right)}$
given by 

\begin{equation}
\Delta r_{g\left(n_{max}-1\rightarrow n_{max}\right)}=2\left(\sqrt{\frac{3}{2}}-1\right)\approx0.449.\label{eq: maximum}
\end{equation}
As a consequence, also the value of the Schwarzschild ``volume quantum''
increases with increasing values of the BH energy level. Hence, the
maximum value of the Schwarzschild ``volume quantum'' is obtained
multiplying $\Delta r_{g\left(n_{max}-1\rightarrow n_{max}\right)}$
by the Bekenstein area quantum, obtaining 
\begin{equation}
\begin{array}{c}
\Delta V_{\left(n_{max}-1\rightarrow n_{max}\right)}=\Delta V_{\left(n_{max}\rightarrow n_{max}-1\right)}=\\
\\
=16\pi\left(\sqrt{\frac{3}{2}}-1\right)\approx2.82.
\end{array}\label{eq: maximum volume quantum}
\end{equation}
It is intriguing and remarkable that both of the results of Eqs. (\ref{eq: maximum})
and (\ref{eq: maximum volume quantum}) do NOT depend on the BH original
mass. In fact, this implies that the behavior of BH evaporation should
be the same for ALL the Schwarzschild BHs when the Planck scale is
approached.

\subsection*{Conclusion remarks}

In this paper we have shown that our recent results in BH quantum
physics {[}4 - 6{]} have the consequence that space-time quantization
could be energy-dependent. Thus, the intriguing result is that, in
a certain sense, space-time could be neither discrete nor continuous.
Our analysis permitted also to show that the ``volume quantum''
of the Schwarzschild space-time increases with increasing energy during
BH evaporation and arrives to a maximum value when the Planck scale
is reached and the GUP prevents the total BH evaporation. As this
result does not depend on the BH original mass, the remarkable consequence
is that the behavior of BH evaporation should be the same for ALL
the Schwarzschild BHs when the Planck scale is approached. 

\subsection*{Acknowledgements }

This study was funded by the Research Institute for Astronomy \& Astrophysics
of Maragha (RIAAM) under research project No. 1/4717-119.

\end{document}